\documentclass[aps,prd,preprint,groupedaddress,superscriptaddress,nofootinbib,eqsecnum]{revtex4-2}
\usepackage[T1]{fontenc}

\usepackage{amssymb}
\usepackage{amsmath}
\usepackage{cancel}
\usepackage{mathtools}
\usepackage{bbm}
\usepackage{faktor}
\usepackage{tensor}
\usepackage{bbm}
\usepackage{shuffle}
\usepackage{slashed}
\usepackage{multirow}

\usepackage{graphicx}
\usepackage[export]{adjustbox}
\usepackage[usenames,dvipsnames]{xcolor}
\usepackage{tikz}
\usetikzlibrary{calc}

\usepackage{relsize}
\usepackage{titlesec}
\usepackage{setspace} 
\usepackage{placeins}
\usepackage{subcaption} 
\usepackage{enumitem}
\usepackage{diagbox}

\usepackage[textsize=tiny]{todonotes}

\definecolor{allOrderBlue}{rgb}{0.4,0.5,1}
\definecolor{patternBlue}{rgb}{0,0,1}
\definecolor{photonRed}{rgb}{1,0.2,0.2}

\def\Ord{\mathcal{O}}

\def\backdexpatch latexntA{\hspace*{-6mm}}
\def\nsand#1.#2.#3{%
  \left\langle\smash{#1}{\vphantom1}\right|{#2}%
  \left|\smash{#3}{\vphantom1}\right]}
\def\nsandaa#1.#2.#3{%
  \left\langle\smash{#1}{\vphantom1}\right|{#2}%
  \left|\smash{#3}{\vphantom1}\right\rangle}
\def\nsandbb#1.#2.#3{%
  \left[\smash{#1}{\vphantom1}\right|{#2}%
  \left|\smash{#3}{\vphantom1}\right]}
\def\nsandba#1.#2.#3{%
  \left[\smash{#1}{\vphantom1}\right|{#2}%
  \left|\smash{#3}{\vphantom1}\right\rangle}

\def\keta#1{%
  \left|\smash{#1}{\vphantom1}\right\rangle}
\def\ketb#1{%
  \left|\smash{#1}{\vphantom1}\right]}
\def\spa#1.#2{\left\langle#1\,#2\right\rangle}
\def\spb#1.#2{\left[#1\,#2\right]}

\def\spash#1.#2{\spa{\smash{#1}}.{\smash{#2}}}
\def\spbsh#1.#2{\spb{\smash{#1}}.{\smash{#2}}}

\def\sect#1{Sect.~{\ref{#1}}}

\def\sects#1#2{Sects.~\ref{#1} and~\ref{#2}}

\def\eqn#1{eq.~(\ref{#1})}

\def\Eqn#1{Eq.~(\ref{#1})}

\def\<{\langle}
\def\>{\rangl\noaffiliatione}

\newbox\charbox
\newbox\slabox
\def\s#1{{      
    \setbox\charbox=\hbox{$#1$}
    \setbox\slabox=\hbox{$/$}
    \dimen\charbox=\ht\slabox
    \advance\dimen\charbox by -\dp\slabox
    \advance\dimen\charbox by -\ht\charbox
    \advance\dimen\charbox by \dp\charbox
    \divide\dimen\charbox by 2
    \raise-\dimen\charbox\hbox to \wd\charbox{\hss/\hss}
    \llap{$#1$}
  }}
\def\cut#1{{      
    \setbox\charbox=\hbox{$#1$}
    \setbox\slabox=\hbox{$|$}
    \dimen\charbox=\ht\slabox
    \advance\dimen\charbox by -\dp\slabox
    \advance\dimen\charbox by -\ht\charbox
    \advance\dimen\charbox by \dp\charbox
    \divide\dimen\charbox by 2
    \raise-\dimen\charbox\hbox to \wd\charbox{\hss$|$\hss}
    \llap{$#1$}
  }}

\DeclareMathOperator{\Tr}{Tr}
\newcommand{\Gram}{G}

\def\oneloop{{(1)}}

\newcommand{\AOne}{A^\oneloop}

\usepackage{orcidlink}

\usepackage[]{hyperref}

\graphicspath{{./figures/}} 

\begin{document}

\newcommand{\Ampl}{\mathcal{A}}
\newcommand{\AmplB}{\mathcal{\bar A}}
\def\Int{\textrm{Int}}

\hfuzz=15 pt

\title{Serendipitous Syzygies of Scattering Amplitudes}

\def\squeezev{\vspace*{-2.5mm}}

\author{David~A.~Kosower\orcidlink{0000-0002-9087-0071}}
\affiliation{\looseness=-1%
  \linespread{1}\selectfont%
  Institut de Physique Th\'eorique, CEA, CNRS, Universit\'e Paris--Saclay,
  F--91191 Gif-sur-Yvette cedex, France
  \\ {\textsf{\rm\sf David.Kosower@ipht.fr}}
}

\author{Sebastian P{\"o}gel\orcidlink{0000-0003-4323-9743}}
\affiliation{\looseness=-1%
  \linespread{1}\selectfont%
 Paul Scherrer Institut, CH--5232 Villigen PSI, Switzerland\\ {\textsf{\rm\sf sebastian.poegel@psi.ch}}
}
\date{May 20, 2025}

\begin{abstract}
We study linear relations between color-ordered all-plus
amplitudes at one loop in Yang--Mills theory.  
We show that on general
grounds, there are $(n-1)!/2-2$ relations for $n\ge 5$, leaving only two independent color-ordered amplitudes.
We present two complementary approaches to finding such relations: one using numerical linear algebra and the other using syzygies in computational algebraic geometry. 
We obtain explicit forms for all relations through $n=7$.  
We also study relations for the tree-level MHV amplitudes through $n=8$. 
The latter relations include the well-known color and
Bern--Carrasco--Johansson identities.
\end{abstract}

\pacs{\hspace{1cm}}

\preprint{1}{PSI-PR-25-05}

\maketitle
\newpage
\thispagestyle{empty}

\section{Introduction}
\label{IntroSection}

\def\Obs{\mathcal{O}}
\def\obs{v}
\def\Pss{P^{**}}

Scattering amplitudes in Yang--Mills theories are functions of 
the colors and momenta of
the external particles.  The amplitudes are most conveniently 
written as sums over products of
color traces and color-ordered or partial amplitudes, separating 
the dependence on the
different kinds of degrees of freedom.  The underlying Lie symmetry then implies linear relations
between different partial amplitudes with constant coefficients. 
The best known of these are the Kleiss--Kuijf (KK)
identities~\cite{Kleiss:1988ne}, which reduce the number
of independent tree-level all-gluon amplitudes to $(n-2)!/2$.
Bern, Carrasco, and Johansson (BCJ) discovered an additional set of identities~\cite{BCJ} for tree-level
partial amplitudes, with coefficients linear in
momentum invariants.  The BCJ relations reduce the
number of independent amplitudes to $(n-3)!$.
Both kinds of relations are all independent of the helicity configuration
of the external particles.

Looking beyond tree level, 
Bern, Dixon, Dunbar, and one of the authors
found group-theory relations for general one-loop
amplitudes. 
Edison and Naculich classified~\cite{Naculich:2011ep} all relations of purely group-theory origin to all loop orders for amplitudes with up to six external gluons. 
Dunbar studied~\cite{Dunbar:2023ayw} relations between 
two-loop all-plus amplitudes, with some all-$n$ results.
At one loop and beyond,
nearly all helicity amplitudes have branch 
cuts, making it more challenging to find
linear relations beyond those implied by group theory.
At one loop, however,
there are two classes of non-trivial amplitudes that are still 
purely rational: the all-plus and the 
single-minus (along with their helicity conjugates).  
A decade ago, Bjerrum-Bohr, Damgaard, Johansson, 
and S{\o}ndergaard~(BBDJS)~\cite{Bjerrum-Bohr:2011jrh}
observed additional relations for these two classes of 
amplitudes.  They motivated the relations
via a diagrammatic representation for these amplitudes.
Some of the BBDJS relations have
constant coefficients, and some have coefficients polynomial 
in the invariants.  More recently, 
Dixon and Morales (DM)~\cite{Dixon:2024mzh}
used an observation by Costello~\cite{Costello:2023vyy} to give
an interpretation of the constant-coefficient BBDJS identities
as having partly a group-theory origin, 
and also to give a proof of completeness.  (Both BBDJS and
DM also discuss the one-minus amplitudes, which contain
multiparticle poles, and thus have a more complicated structure
of relations.)

In this paper, we explore such relations more fully,
focusing on those with non-constant coefficients.  
We show in \sect{GeneralSyzygySection} that they must exist
in general, by constructing a related set of
functions with the same relations.  
We compute the full sets of relations
for the five-, six-, and seven-point
amplitudes, and isolate generating vectors for the
five- and six-point relations.  
We present two
different approaches to systematically searching for such relations.  One is based on numerical linear algebra,
presented in \sect{NumericalSyzygiesSection};
the other uses factorization and
computational algebraic geometry, and
is presented in \sects{FactorizationSection}{GeneratorSection}.
We present results for the all-plus amplitudes in
\sect{AllPlusSection}, and for the tree-level MHV amplitudes in
\sect{MHVSection}.  We record an interesting observation
about the all-plus relations in \sect{BoseSymmetryObservation}.
We summarize in \sect{ConclusionsSection}.

\section{Color Decomposition}
\label{NotationSection}

In Yang--Mills theories, we can write amplitudes in a color-decomposed
form,
\begin{equation}
    \mathcal{A}^{(L)}_n=\sum_{c=1}^{d_n(L)}A_{n;c}^{(L)}t_c^{(L)}.
\label{ColorDecomposition}
\end{equation}
where $d_n(L)$ is the dimension of the trace basis at $L$ loops for $n$ particles.  We focus on pure gluon amplitudes with $SU(N)$ color
group, but the approach is general.  We adopt the notation of
Edison and Naculich~\cite{Naculich:2011ep}.  We will be 
interested in
linear relations between the color-ordered amplitudes 
$A_{n;c}$.

Each element $t_c$ of the trace basis has the form,
\begin{equation}
\label{eq:color_trace_element}
    N^k \bigl[\Tr(\alpha_1)\Tr(\alpha_2)\ldots\Tr(\alpha_m)+
    (-1)^n 
    \Tr(\alpha_1^T)\Tr(\alpha_2^T)\ldots\Tr(\alpha_m^T)\bigr]\,,
\end{equation}
where $m$ is the number of color traces and $k$ denotes the number of explicit factors of $N$ (which can also be interpreted as traces containing the identity). 
The $\alpha_i$ are products 
of $\mathrm{SU}(N)$ generators, which together form a partition of 
all generators 
corresponding to the external particles.
At two loops or beyond, there are subleading-color contributions
with fewer traces, so generally $k \ne L-m+1$.
The color-ordered amplitude $A^{(L)}_{n;c}$ corresponding to 
$t_c$ can also be labeled as follows,
\begin{equation}
   A^{(L)}_{n;c}= A^{(L,k)}_{n;n_1;\ldots;n_m}(\alpha_1;\alpha_2;\ldots;\alpha_m),
\end{equation}
where $n_i=|\alpha_i|$.
These amplitudes are invariant under exchange of any two
of the sets $\alpha_i$, as well as the reflection operation, 
\begin{equation}
    A^{(L,k)}_{n;n_1;\ldots;n_m}(\alpha_1;\alpha_2;\ldots;\alpha_m)=(-1)^n A^{(L,k)}_{n;n_1;\ldots;n_m}(\alpha_1^T;\alpha_2^T;\ldots;\alpha_m^T),
\end{equation}
justifying the combination of both cyclic and anti-cyclic color 
traces in the basis elements in \eqn{eq:color_trace_element}.
In addition, each element of the trace basis and associated 
amplitude are invariant under cyclic permutations of each of the 
$\alpha_i$.  At one loop, we have leading-color
terms, with single-trace
color factors (and a corresponding factor of $N$); and
subleading-color terms, with double-trace color factors.
The subleading-color amplitudes, coefficients of the
latter in \eqn{ColorDecomposition}, are related
by color algebra relations~\cite{Bern:1994zx}
to the leading-color ones.  
Accordingly, in this article we shall consider only 
relations between leading-color
amplitudes.  Both at tree level and at one loop,
cyclicity and (anti-)reflection properties of the color
traces leave us with $(n-1)!/2$ distinct permutations of
the amplitude arguments to consider.
We take this set to be the permutations $\sigma$ 
on legs $2,\ldots,n$ with $\sigma_2<\sigma_n$.
To organize our calculations and results, let us order 
these permutations lexically.  

\section{Syzygies of Rational Functions}
\label{GeneralSyzygySection}

\def\PolyName{P}
\def\PolyId#1#2{\PolyName^{[#1]}_{#2}}
\def\PolyKName{{\mkern 3mu\widetilde{\mkern-3mu \PolyName\mkern-2.5mu}\mkern 2.5mu}}
\def\PolyK#1#2{\PolyKName^{[#1]}_{#2}}
\def\permset#1{S^*_{#1}}
\def\Vdim{n_A}
We are interested in the existence of relations,
\begin{equation}
\sum_{\sigma\in \permset{n}} 
\PolyId{r}{n}[\sigma] A_n(\sigma) = 0\,,
\label{GeneralSyzygy}
\end{equation}
where the $\PolyId{r}{n}[\sigma]$ are polynomials of 
degree $r$ in the Mandelstam
invariants of the external particles,
so that we may call these relations \emph{syzygies};
and where $\permset{n}$ is the set of $\Vdim=(n-1)!/2$ 
distinct permutations.
We fix the helicities of the external particles.

Our definition includes the
known identities originating in the color algebra.
In general, the polynomials for different permutations
$\sigma$ will be \textit{different\/} functions of the invariants, and not merely the same function with arguments related by a relabeling of 
invariants' indices.  This is similar to the case of the
BCJ relations.  Using the (lexical) ordering
of the permutations, we can also write these relations
as vectors of dimension $\Vdim$.

\def\lv{\varepsilon}
\def\Rat{\mathop{\rm Rat}\nolimits}
\def\Poly{\mathop{\rm Poly}\nolimits}
\def\unia{\vphantom{ijkl}}
\def\unib{\vphantom{\mu\nu\lambda\rho}}
How many identities should we expect?  Consider,
for example, the set
of $(n-1)!/2$ all-plus amplitudes~\cite{AllPlus}
given by permutations of,
\begin{equation}
A_{n;1}(1,2,\ldots,n) = \frac{i}{48\pi^2}
\frac{H_n}{\spa1.2\spa2.3\cdots\spa{(n\!-\!1)}.n\spa{n}.1}\,,
\end{equation}
where 
\begin{equation}
H_n = -\hspace*{-4mm}\sum_{1\le i_1<i_2<i_3<i_4\le n}
\hspace*{-4mm}
\spa{i_1}.{i_2}\spb{i_2}.{i_3}\spa{i_3}.{i_4}\spb{i_4}.{i_1}\,.
\end{equation}
We can freely multiply
each element of the set by the same function, and retain
the same identities.  Start by multiplying by the common
spinor denominator; this is the product of angle products of
all pairs,
\begin{equation}
\prod_{i<j}^n \spa{i}.{j}
\end{equation}
The expressions we are considering are now polynomials in
the spinor products.  We can multiply them by a factor chosen
to remove the spinor phase weight, such as,
\begin{equation}
\frac1{\spb1.2\spb2.3\cdots\spb{(n\!-\!1)}.n\spb{n}.1}
\prod_{i<j}^n \spb{i}.{j}\,,
\end{equation}
which is also a polynomial in the spinor products.  The resulting
expressions have no spinor phase weight, and accordingly can be
re-expressed in terms of invariants $s_{ij}$ and 
contracted Levi--Civita
tensors $\epsilon_{ijkl}=\epsilon_{\mu\nu\lambda\rho} 
p_{i\unia}^{\mu\unib}p_{j\unia}^{\nu\unib}
p_{k\unia}^{\lambda\unib}p_{l\unia}^{\rho\unib}$ by organizing the spinors into
traces, and reducing those traces.  

Next, as we are working in four spacetime dimensions,
when we have five or more 
external particles, we can
choose a basis of four momenta (for example, 
$k_1, k_2, k_3, k_4$).  We can express every 
contracted Levi--Civita
tensor in terms of that of the basis vectors 
($\epsilon_{1234}$ for the example)
by expanding its momenta in terms of the basis.  The coefficients
of that basis are ratios of Gram determinants, which in turn
are polynomials in the invariants.   
The Gram determinant $\Gram$ is defined by,
\begin{equation}
G\bigg(\genfrac{}{}{0pt}{}{p_1,\ldots,p_m}{q_1,\ldots,q_m}\bigg) =
 \det_{i,j}\big(2p_i\cdot q_j\big)\,.
\end{equation}
(When both rows are identical, we write a single list of arguments.)

The coefficients' denominator
is the Gram determinant of the basis vectors, $G(1\,2\,3\,4)$.
Multiply by up to four powers of this Gram determinant to clear
the newly introduced denominators.  We finally have a set
of polynomial expressions, each in the form,
\def\PolyA{Q^{[0]}}
\def\PolyB{Q^{[\epsilon]}}
\begin{equation}
\Poly_{a}(\{s_{ij}\}) + \Poly_{b}(\{s_{ij}\})\,\lv_b\,,
\end{equation}
which have exactly the same identities as the original set
of amplitudes.  (In this expression, $\lv_b$ is the
Levi--Civita tensor of the basis momenta.)
It is now easy to see that any triplet of
expressions, and hence any triplet of amplitudes 
$(A_1,A_2,A_3)$, must have
a polynomial identity.  We can see this by constructing it
explicitly.  Define,
\begin{equation}
\begin{aligned}
\xi_1 &= \Poly_{3,a} \Poly_{2,b} - \Poly_{2,a} \Poly_{3,b}\,,
\\ \xi_2 &= \Poly_{1,a} \Poly_{3,b} - \Poly_{3,a} \Poly_{1,b}\,,
\\ \xi_3 &= \Poly_{2,a} \Poly_{1,b} - \Poly_{1,a} \Poly_{2,b}\,.
\end{aligned}
\end{equation}
Then $(\xi_1,\xi_2,\xi_3)$ is a syzygy, that is,
\begin{equation}
\xi_1 A_1 + \xi_2 A_2 + \xi_3 A_3 = 0\,.
\label{ThreeSyzygy}
\end{equation}

Each triplet of amplitudes will have an identity, and 
$m$ amplitudes
will have $m-2$ identities, so that we have two independent
amplitudes.  In the four-point case, the Levi--Civita tensor
vanishes identically, so that we have two identities for
the three independent amplitudes, and one independent amplitude.

The \emph{existence\/} of syzygies for rational
amplitudes is thus not surprising.  Indeed, as the above argument
shows, any collection of rational functions of spinors with the
same spinor phase weight will exhibit such relations.
For generic functions of the
same engineering dimensions of numerators and denominators
as amplitudes, we would expect the $\Poly_{i,x}$ to be of degree
$\Ord(n)$ (or higher), and hence the polynomials in 
a relation~\eqref{GeneralSyzygy}
to be of degree $\Ord(n^2)$ or higher. 
What \emph{is\/} surprising is 
the low
polynomial degree at which they arise, and at least for the 
all-plus amplitude, the
simple structure of the coefficients.   
As already shown by BBDJS, there
are relations of degree zero and two for the all-plus amplitudes.
For this specific helicity amplitude these relations 
turn out to exhaust all possible relations~\cite{Dixon:2024mzh}.  

It is possible to search for syzygies by parametrizing them
order by order in the degree of invariants, and solving for coefficients
numerically (or showing that no solution exists).  We discuss
this brute-force approach in the next section.  What the general
argument above fails to consider are the special properties
of amplitudes, in particular factorization.  We discuss how
to derive equations relying on factorization in 
\sect{FactorizationSection}, and how to solve the resulting
equations in subsequent sections.

\section{Finding Syzygies Numerically}
\label{NumericalSyzygiesSection}

\def\fflas{\textsf{FFLAS--FFPACK}}
\def\rank{\mathop{\rm rank}\nolimits}
The simplest approach to finding relations between a 
given vector of amplitudes is via linear algebra performed numerically.
For example, to find constant (degree-zero) relations, we can evaluate the vector of $\Vdim$ amplitudes on $\Vdim$ different momentum configurations, 
obtaining a $\Vdim\times \Vdim$ matrix $V$ of values. 
It is important to do so using exact arithmetic --- that
is, either rational or finite-field momentum configurations --- to avoid errors due to floating-point arithmetic in 
large matrices.
It is straightforward to generate (complex) rational momentum 
configurations using the momentum-twistor
approach described in appendix E of 
ref.~\cite{Kosower:2022bfv}.
As we have sampled the vector of amplitudes $\Vdim$ times, the 
only non-trivial relations of the numeric vectors are those 
inherited from the original amplitudes. 
Thus we can recover the set of degree-zero relations by simply 
computing the null space $\ker{V}$.
The number of linearly independent relations between the 
vector of amplitudes is given by $\dim \ker V=\dim V-\rank V$. 

\def\Ahat{\widehat A}
The momentum twistor parameterization of 
ref.~\cite{Kosower:2022bfv} represents the space of $n$ 
massless on-shell momenta in a momentum-conserving configuration
using $3n-10$ parameters $z_i$.
The parametrization implicitly fixes the freedom associated 
with little group transformations, the spinor phase weights,
in an arbitrary way.
As the amplitudes have a non-zero spinor phase weight, each 
numerical sample of the vector will have a different phase 
associated to it. 
To obtain relations from numerical evaluations, all evaluations 
should have the same phase associated to them.
We can achieve this by multiplying all amplitudes by the
same overall factor to absorb this phase information.
We define the adjusted vector of amplitudes,
\begin{equation}
\bigl(\Ahat_n(\sigma^{(1)}),\,\Ahat_n(\sigma^{(2)}),\,
\ldots,\Ahat_n(\sigma^{(\Vdim)})\bigr) =
\phi(1,\ldots,n) \bigl(A_n(\sigma^{(1)}),\, A_n(\sigma^{(2)}),\,
\ldots, A_n(\sigma^{(\Vdim)})\bigr)\,,
\end{equation}
where the function $\phi$ is chosen such that the $\Ahat$s
are free of spinor phase weight.
In the case of all-plus amplitudes we can choose, 
\begin{equation}
	\phi(1^+,\ldots,n^+)=\spa{1}.2\spa{2}.3 \ldots \spa{n}.1,
\end{equation} 
while for MHV amplitudes with the two negative helicities 
$m_{1,2}$ and remaining indices $I_+$, we can choose
\begin{equation}
\phi(1,\ldots n)=
    \spb{m_1}.{m_2}^2 
    \prod_{i\in I_+} \spa{i}.{\textrm{succ}(i)}\,.
\end{equation}
($\textrm{succ}(i)$ is the succeeding element of $I_+$,
understood cyclicly.)
As we are modifying all amplitudes by the same overall factor,
any relation we obtain for the vector of $\Ahat$s is also
a relation for the original vector of 
amplitudes.

Amplitudes over real kinematics are typically complex.
Using a parametrization based on momentum twistors 
allows us to restrict the values of amplitudes.
In particular, amplitudes which are rational functions of
spinors, like those we are considering here,
will take on rational 
values\footnote{We divide out an overall factor of $i$.} 
for rational values of the
parameters $z_i$: $A_n(z_i)\in \mathbb{K}$ when
$z_i\in \mathbb{K}$ for some field $\mathbb{K}$.
To generate the matrix $V\in\mathbb{K}^{\Vdim\times\Vdim}$, 
we evaluate the vector of $\Ahat$s on $\Vdim$ random sets of 
$\{z_1,\ldots,z_{3n-10}\}\in \mathbb{K}^{3n-10}$.
If we are interested in obtaining the explicit expressions for 
the relations and the dimension $\Vdim$ is 
small---up to $\Ord(2000)$---we can use $\mathbb{K}=\mathbb{Q}$.
If we are instead interested just in counting the number of relations or are dealing with larger dimensions,
say $\Vdim$ of $\Ord(10\,000)$, we instead use a finite field $\mathbb{K}=\mathbb{F}_p$ for some suitably large prime for greater computational efficiency.
In the latter case we would have to reconstruct the explicit (rational) relations from evaluations over a sufficient number of different prime fields $\mathbb{F}_p$.
To obtain the null space of the matrix $V$ over such a finite 
field, we use the $\texttt{C++}$ library \fflas~\cite{fflas}, 
which provides parallelized
functions for modular linear algebra operations, and performs significantly better than the corresponding \textsl{Mathematica} built-in functions.%
\footnote{Due to technical constraints, the library is limited to finite fields with $32$-bit primes.
However, we observed significantly reduced efficiency close to this limit, consistent with the expected increase in modular operations.
For this reason we perform computations over fields with $p = 2^{24}-3={16777213}$ or nearby primes, which provides a balance between performance and a sufficiently sized prime required for reconstructing rational numbers from the finite field values.
}

We also want to go beyond degree-zero relations, and allow for 
factors of Mandelstam invariants multiplying the color-ordered
amplitudes.
We take the usual invariants,
\begin{equation}
s_{i_1 \ldots i_k} = \Bigl(\sum_{j=1}^{k}p_{i_j}\Bigr)^2\,,
\end{equation}
and consider 
the ring of polynomials in the invariants with coefficients over
the field $\mathbb{K}$ (which will again be either $\mathbb{Q}$
or $\mathbb{F}_p$).  
Momentum conservation and Gram-determinant constraints,
\begin{equation}
    G\begin{pmatrix}
         i_1& i_2& i_3 & i_4 & i_5\\
         j_1& j_2& j_3 & j_4 & j_5\\    
    \end{pmatrix}=0\,,
\end{equation}
reflecting the four-dimensionality of momenta, lead to relations between polynomials in this ring.  (We have abbreviated
the notation here by using the indices of the momenta as
arguments to the Gram determinant.)

\def\MonomialBasis{\mathcal{B}}
In order to ensure that the relations we find are indeed 
syzygies of the amplitudes and not a result of
identities between invariants, we should establish
a linearly independent basis $\MonomialBasis^{(n)}_m$ of 
monomials of any given degree $m$.
For general $n$, a degree $m=1$ basis is given by,
\begin{equation}
    \MonomialBasis^{(n)}_1 = 
    \{s_{12},\mathellipsis,s_{n1}\}\cup\bigcup_{i=2}^{n-4}\{s_{1,\mathellipsis ,1+i},
    \mathellipsis,s_{n-1-i,\mathellipsis, n-1}\}.
\end{equation}
This basis has
\begin{equation}
    \bigl|\MonomialBasis^{(n)}_1\bigr|=n+\sum_{i=2}^{n-4}n-i-1=\frac{n(n-3)}{2}
\end{equation}
elements.  This basis is valid in general spacetime dimension,
and hence has more degrees of freedom than the usual
$3n-10$ of a four-dimensional momentum-conserving configuration of
$n$ massless momenta.  The difference is
accounted for by the $\tfrac{1}{2} (n-4)(n-5)$ Gram determinant 
relations.  These relations are of degree five, and so
are irrelevant for any analysis with $m\le 4$.

For monomials of degree $m$, we can generate a basis $\MonomialBasis^{(n)}_m$ by taking symmetric products of the corresponding $m=1$ basis,
\begin{equation}
\MonomialBasis^{(n)}_m=\left\{
    s_{1} \cdots s_{m}|s_i\in \MonomialBasis^{(n)}_1
\right\}\,,
\end{equation}
which gives us 
\begin{equation}
    \bigl|\MonomialBasis^{(n)}_m\bigr| =
    \frac{(\bigl|\MonomialBasis^{(n)}_1\bigr|)_{m}}{m!} =
    \frac{1}{m!}\left( \frac{n(n-3)}{2} \right)_{m}
\end{equation}
unique monomials.
Here \((x)_n=x (x+1) \mathellipsis (x+n-1)\) is the Pochhammer 
symbol.
For $n\ge 6$ and $m\ge 5$ this basis of monomials is overcomplete,
and requires the imposition of the Gram-determinant relations.
Once this is done, we are left with a reduced basis, which we 
denote by $\overline{\MonomialBasis}^{(n)}_m$.
Just as for the amplitudes, we can evaluate these bases of 
monomials in terms of the same momentum twistor parameters 
already used for the vector of amplitudes.

\def\tensorvec{T}
In order to obtain syzygies with polynomial coefficients
of degree $m$ numerically, we form the
tensor product of the vector of 
adjusted amplitudes and the basis of degree-$m$ invariants,
\begin{equation}
    \tensorvec_m = \overline{\MonomialBasis}^{(n)}_m\otimes 
    \bigl(\Ahat_n(\sigma^{(1)}),\,\Ahat_n(\sigma^{(2)}),\,
    \ldots,\Ahat_n(\sigma^{(\Vdim)})\bigr)\,,
\end{equation}
of dimension ${\Vdim}_m=\Vdim\times |\overline{\MonomialBasis}^{(n)}_m|$.
We then proceed as before, evaluating $\tensorvec_{m}$ over
${\Vdim}_m$ numerical samples to obtain a sample matrix $V$.
We then find the null space $\ker{V}$ to obtain the set of
relations between the set of amplitudes times invariants. 
If we are only interested in the relations themselves we can stop here.
However, we would ordinarily want to know the number of 
linearly independent relations as well. 
This number does \textit{not\/} correspond to the matrix rank 
of $V$ as just computed.
Instead, we have to turn the null space $\ker{V}$ into 
relations between the vector of $\Ahat$s,
\begin{equation}
	(\ker{V})\cdot\tensorvec_{m}= M \cdot
    \bigl(\Ahat_n(\sigma^{(1)}),\,\Ahat_n(\sigma^{(2)}),\,
    \ldots,\Ahat_n(\sigma^{(\Vdim)})\bigr)\,,
\end{equation}
where the matrix elements of $M$ are now polynomials of
degree $m$ in the invariants.
The rank of $M$ then gives us the
number of independent syzygies with coefficients of degree $m$.
These syzygies fall into two categories. 
There are relations which are genuinely of degree $m$,
and cannot be expressed as linear combinations of lower-degree
syzygies.
The matrix $M$ also contains all syzygies of lower degree,
multiplied by polynomials raising their degree to $m$.
We can call the latter \textit{inherited\/} syzygies.
We can separate the inherited from the true relations by computing the quotient space of degree $m$ relations by all relations with degree $m'<m$.

The numerical approach just described has some obvious 
limitations.  It has diminishing performance in both runtime 
and especially in memory usage with increasing dimension 
${\Vdim}_m$, scaling roughly as $\Ord({\Vdim}_m^3)$ and 
$\Ord({\Vdim}_m^2)$ respectively because of the underlying linear 
algebra.
In practice, using $\mathbb{K}=\mathbb{F}_{16777213}$ we are able to work out cases with ${\Vdim}_m \sim \Ord(100\,000)$.
Furthermore, the resulting relations are generally unstructured, as they are semi-randomly chosen through
the null space computation, 
with additional mixing of inherited and true relations. 
However, a significant benefit lies in its simplicity: very 
little additional information is required aside from deriving a 
suitable rational function representation of the underlying 
amplitudes and invariants, at which point it is agnostic 
to any additional structure.
This makes it very versatile in quickly testing for relations for 
different sets of (rational) amplitudes, such as tree-level MHV
as well as one-loop all-plus amplitudes.
Furthermore, it is well-suited to obtaining the number of 
independent amplitudes at any specified polynomial
degree in the invariants (or equivalently the number of 
independent relations).
We display the counting of syzygies and of independent amplitudes 
for the one-loop all-plus and the tree-level MHV at various degrees in 
Tables~\ref{tab:one-loop_relations_summary} 
and~\ref{tab:tree_MHV_relations_summary}.
Curiously, we observe that in both cases a sufficient number of relations quadratic in Mandelstams exist to reduce all partial amplitudes to just two independent ones, i.e.~the minimal number allowed.
In the case of the all-plus we have verified this up to seven and in the MHV case up to eight points.
In all cases we have exact --- i.e.~$\mathbb{K}=\mathbb{Q}$ --- expressions for these relations.
We provide them in the form of generators over permutations, as discussed in section~\ref{GeneratorSection}.

\begin{table}
        \begin{subtable}{0.49\textwidth}
            \centering
            \begin{tabular}{c|c|c|c|c|c|c|c}
             \multirow{2}{*}{\backslashbox{\hspace*{10mm}\centering $n$}{degree}}& 
                 & \multicolumn{2}{c|}{0} & 
                \multicolumn{2}{c|}{1} & \multicolumn{2}{c}{2}\\ 
                & $A$ & syz&$A$& syz&$A$& syz&$A$ \\ \hline
                4 & 3 & 2 & 1 & -- & 1 & -- & 1 \\
                5 & 12 & 6 & 6 & -- & 6 & 4 & 2 \\
                6 & 60 & 25 & 35 & -- & 35 & 33 & 2\\
                7 & 360 & 135 & 225 & -- & 225 & 223 & 2\\
            \end{tabular}
        \end{subtable}
        \caption{\label{tab:one-loop_relations_summary} 
        Number of syzygies of different degrees (syz) 
        and of resulting 
        independent amplitudes (A) after imposition of all
        syzygies up to the given degree, 
        for the one-loop all-plus 
        amplitude.}
\end{table}

\begin{table}
        \centering
        \begin{subtable}{0.5\textwidth}
            \centering
            \begin{tabular}{c|c|c|c|c|c|c|c}
             \multirow{2}{*}{\backslashbox{\hspace*{10mm}\centering $n$}{degree}}& 
                 & \multicolumn{2}{c|}{0} & 
                \multicolumn{2}{c|}{1} & \multicolumn{2}{c}{2}\\ 
                & $A$ & syz&$A$& syz&$A$& syz&$A$ \\ \hline
                4 & 3 & 1 & 2 & 1 & 1 & -- & 1  \\
                5 & 12 & 6 & 6 & 4 & 2 & -- & 2  \\
                6 & 60 & 36 & 24 & 18 & 6 & 4 & 2 \\
                7 & 360 & 240 & 120 & 96 & 24 & 22 & 2\\
                8 & 2520 & 1800	 & 720 & 600 & 120 &118 &2\\
            \end{tabular}
        \end{subtable}%
        \caption{\label{tab:tree_MHV_relations_summary} 
        Number of syzygies of different degrees (syz) 
        and of resulting 
        independent amplitudes (A) after imposition of all
        syzygies up to the given degree, 
        for the tree-level MHV amplitude.  
        The `0' column corresponds
        to the Kleiss--Kuijff relations, while the `1' column
        corresponds to the BCJ identities.
        }
\end{table}

\section{Syzygies from Factorization}
\label{FactorizationSection}

\def\Res{\mathop{\rm Res}}
\def\filter{F}
\def\unfilter{{\mkern 4mu\overline{\mkern-4mu F\mkern-1mu}\mkern 1mu}}

The amplitudes in the sum in \eqn{GeneralSyzygy} will have poles
in the invariants.  On each pole, the amplitudes
that have a pole will factorize into a product of lower-point
amplitudes.  For the all-plus amplitude, poles arise only
at spinor invariants $\spa{i}.{j}$.  Examine the residue
at such a pole,
\begin{equation}
\sum_{\sigma\in \permset{n}}\Res_{\spa{i}.{j}=0}
\PolyId{r}{n}[\sigma]\, A_n(\sigma) = 0\,.
\label{ResidueI}
\end{equation}
(The set $\permset{n}$ is defined below \eqn{GeneralSyzygy}.)
A given amplitude $A(\sigma)$ will have a pole at $\spa{i}.{j}$
only if $i$ and $j$ are (cyclicly) adjacent in the permutation
$\sigma$.  Denote this by $(ij)\in \sigma$; and
denote the subset of a set of permutations $S$ which contain
$ij$ cyclicly adjacent by $\filter_{ij}(S)$,
\begin{equation}
\filter_{ij}(S) = \big\{ \rho \in S\,|\, (ij)\in \rho\big\}\,.
\end{equation}
Amplitudes without a pole will have zero residue.

\def\permred#1{\breve{#1}}

Taking the residue in \eqn{ResidueI}, and dropping overall
factors, we obtain,
\begin{equation}
\sum_{\sigma\in B_{ij}} \PolyId{r}{n}[\sigma] \,\tau^{(ij)}_{\sigma} 
\omega^{(ij)}_{\sigma}
A_{n-1}(\permred\sigma) A_3((-P)^- i^+ j^+) = 0\,.
\label{PoleSyzygyEquation}
\end{equation}
In this equation, $B_{ij}=F_{ij}(\permset{n})$;
$\permred\sigma$ denotes the reduction 
of $\sigma$ to a permutation of the $n-1$ symbols
$\{1,2,i\!-\!1,P,j\!+\!1,\ldots,n\}$ obtained by replacing 
`$ij$' by `$P$' and using cyclic symmetry and reflection
(anti-)symmetry to put it into canonical form; 
$\tau^{(ij)}_{\sigma}$ denotes the sign that
may arise from putting the three-point amplitude into the given 
form from the original ordering
of $ij$ in $\sigma$; and $\omega^{(ij)}_{\sigma}$ denotes the 
sign that may arise from 
reducing $\sigma$ to $\permred\sigma$ and putting the latter
into canonical form.  (Recall that we use a canonical form 
for $\sigma$ 
where $1$ always appears first, and where $\sigma_2<\sigma_n$.)

To lighten the notation, define
\begin{equation}
\zeta^{(ij)}_\sigma = \tau^{(ij)}_\sigma \omega^{(ij)}_\sigma\,.
\end{equation}

The factor of $A_3((-P)^- i^+ j^+)$ is common to all terms, 
and non-vanishing for generic momenta, so we can divide it out.
We are left with the following equation, 
\begin{equation}
\sum_{\sigma\in B_{ij}} 
\zeta^{(ij)}_{\sigma}\,\PolyId{r}{n}[\sigma] 
\,A_{n-1}(\permred\sigma) = 0\,.
\label{ReducedSyzygyGeneralPre}
\end{equation}
In general, there are two syzygy entries $\PolyId{r}{n}$ 
that appear in the 
coefficient of each $(n-1)$-point amplitude
$A_{n-1}(\permred\sigma)$, so that we can rewrite
our equation as,
\begin{equation}
\sum_{\xi\in \permred B_{ij}} \Bigl(\sum_{\sigma|\permred\sigma=\xi}
\zeta^{(ij)}_{\sigma}\,\PolyId{r}{n}[\sigma] \Bigr)
A_{n-1}(\xi) =0\,,
\label{ReducedSyzygyGeneral}
\end{equation}
where we denote the set of $\permred\sigma$ corresponding to 
$\sigma\in B_{ij}$ by $\permred B_{ij}$.

As we are proceeding recursively, we assume that we know all 
syzygies for the $(n-1)$-point
amplitudes.  That is, we know all the solutions 
to \eqn{GeneralSyzygy} for $(n-1)$-point amplitudes.
These solutions are expressed in terms of $(n-1)$-point
invariants.
\Eqn{ReducedSyzygyGeneral} tells us that on the pole,
the given sum of 
$\PolyId{r}{n}$s must be
a sum of $(n-1)$-point syzygies, 
\begin{equation}
\sum_{\sigma|\permred\sigma=\xi} 
\zeta^{(ij)}_{\sigma}
\PolyId{r}{n}[\sigma]\bigg|_{\spa{i}.j\, \textrm{pole}} = 
\sum_{(t,v)\in Z_{ij}} f^{(ij)}_{t,v}
\PolyK{t}{n-1|v}[\xi]\,,
\qquad \forall\xi\in \permred B_{ij}\,,
\label{BasicEquationPole}
\end{equation}
where $Z_{ij}$ denotes
the set of indices $(r,v)$ of the known syzygies $\PolyKName$ 
of the $(n-1)$-point amplitudes ($t\le r$).  
This equation represents a set of equations indexed by 
$\xi\in \permred B_{ij}$, as the polynomials come
in sets corresponding to a set of scattering amplitudes.  
This is reflected
in the fact that the coefficients $f_{t,v}$ are \textsl{independent\/}
of $\xi$.  

We seek the solution to the simultaneous sets of equations
for all $ij$ (of which there are of course $n(n-1)/2$ choices
at $n$ points).  We are interested in solutions in terms of
$n$-point invariants, away from the poles as well as on
the poles.

In order to solve the equations, we must first extend the equations
to general kinematics.  This poses the first of several
subtleties we must confront.  We can promote the $(n-1)$-point
invariants to $n$-point invariants by replacing the momentum
$k_P$ by the sum $k_i+k_j$; but we must also allow for
terms that vanish on the pole and hence do not contribute
to the residue.  Given a set of polynomials $Y_{ij}$ which
vanish on the pole, we could extend \eqn{BasicEquationPole}
by adding such polynomials,
\begin{equation}
\sum_{\sigma|\permred\sigma=\xi} 
\zeta^{(ij)}_{\sigma}
\PolyId{r}{n}[\sigma] = 
\sum_{(t,v)\in Z_{ij}} f^{(ij)}_{t,v}
\PolyK{t}{n-1|v}[\xi]
+ \sum_{g\in G_{ij}}
c_g^{(ij)}[\xi]\, g\,,\qquad \forall\xi\in \permred B_{ij}\,,
\label{BasicEquation}
\end{equation}
where $G_{ij}$ is the Gr{\"o}bner basis of $Y_{ij}$, which
gives the complete set of independent polynomials we could add.

The relevant kinematics are \emph{complex\/}
(or holomorphic) collinear kinematics, 
with $\keta{i}\parallel\keta{j}$,
rather than ordinary collinear kinematics, with 
$k_i\parallel k_j$ with $\ketb{i}\parallel\ketb{j}$ as well.
We do not require the invariants to
be real, and the computational algebraic geometry tools we shall
employ implicitly take them to be complex.  

The only invariant which vanishes in the complex collinear
kinematics we are using is the pole invariant $s_{ij}$.  
This is problematic, because it does not fully parametrize
the space of (spinorial) functions which vanish on the pole.
As a result, using $G_{ij} = Y_{ij} = \{s_{ij}\}$ will not
give us a complete set of solutions.  As we discuss
in the next section, this is not a problem as long as we
obtain at least a few solutions.  As we explain there,
we can generate all solutions from a limited number of
generators.

An alternative would be to extend the set of polynomials
to include all those which vanish for ordinary collinear
kinematics.  In addition to $s_{ij}$, this set of polynomials
consists of Gram determinants,
\begin{equation}
\Gram\bigg(\genfrac{}{}{0pt}{}{k_i\,k_j}{k_a\,k_b}\bigg)\,,
\end{equation}
for all pairs $(a,b)$.  These are not all independent, however
that lack of independence is taken care of by computing
the Gr{\"o}bner basis.  If we use this extended set,
we now parametrize the required kinematics away from the pole,
at the price of more complicated equations and the possibility
of spurious `solutions' which we would need to prune later.
If we are considering solutions of degree five or higher,
we must also include Gram determinant constraints enforcing
the four-dimensionality of momenta.

The set of equations we would have in this case would have the
form of syzygy equations if the coefficients $f$ and $c$
were themselves polynomial.  This turns out to be true
for some $n$-point syzygies, but not necessarily for all.
Were we to take $f$ and $c$ to be polynomial, we would
still obtain solutions, but they would not all be of minimal
degree.  For the $f$ coefficients, we can proceed as follows.
The coefficients $f_{(0,v)}$ of degree-zero syzygies of 
the $(n-1)$-point amplitudes necessarily appear with constant
coefficients, and we can simply solve for them to eliminate
them.  This would anyway be good practice, as it would simplify
the set of equations.  For the coefficients $f_{(t>0,v)}$,
we can eliminate them, again using tools from computational
algebraic geometry (as implemented, for example, in
the \textsf{Singular\/} function \textsf{eliminate}).  
This elimination increases the number of lower-degree solutions we
obtain, but at the cost of increasing the
computational complexity of solving the equations, because
the polynomial coefficients of the resulting equations are
typically of much higher degree.  The $c$ coefficients cannot 
necessarily be eliminated, however. They start out each 
appearing only once, and 
their elimination would swallow equations.

The equations either before or
after elimination of the $f$ coefficients
can be solved using standard computational algebraic
geometry tools, for example using \textsf{Singular}'s 
\textsf{syz} function.  This yields a set of
candidate solutions.  We still need to impose the original
equation~(\ref{GeneralSyzygy}) on the pole at infinity (if any).  
While one can do this analytically by using a 
BCFW shift~\cite{Britto:2005fq}
and taking the large-parameter limit, it is easiest
and most practical to do this numerically.  The number
of candidate solutions is typically between the number
of final solutions and a small multiple thereof, so that
the size of numerical matrices one must handle is 
much more modest than in the purely numerical approach
of \sect{NumericalSyzygiesSection}.

A somewhat different approach to solving the equations solves
them for a subset of syzygy-vector components (`dependent')
in terms
of other components (`independent')
and terms proportional to $f$ and $c$
functions.  These solutions are then supplemented by
constraint equations involving only invariants alongside
$f$ and $c$ functions.  The independent syzygies parametrize
the different degree-zero solutions.  We can obtain them
by setting all $f$ and $c$ functions to zero.  We could
again treat the constraint equations as syzygy equations,
solving for $f$ and $c$ polynomials.  It turns out that
in some cases, if we simply solve the constraint equations
for a subset of $c$ functions, we obtain rational solutions for
these functions, but the equations for the syzygy-vector
components remain polynomial after substitution of the
values for the $c$s.  When this holds, the remaining
$c$ and $f$ functions parametrize vectors whose linear
combinations (with the $c$s themselves taken to be linear
in the invariants) are true syzygies.

All the solution approaches above leave 
the question of determining whether the
solutions obtained are in fact of minimal degree or
not.  In the case of the all-plus amplitudes, we find
that all solutions (through seven points) are either
of degree zero or degree two. Hence we expect only
degree-two solutions beyond the set of degree-zero
syzygies found by BBDJS and DM.  It is straightforward
to rule out degree-one solutions numerically, showing
that the degree-two ones are in fact of minimal degree.
More generally, one could in principle perform a so-called
\emph{primary decomposition\/} to either establish 
minimality, or to reduce solutions to those of minimal
degree.  In practice, however, available algorithms
(for example, \textsf{Singular}'s \textsf{GTZmod})
are too slow to use for reduction, though they may be
adequate to prove minimality.

\section{Solution Generators}
\label{GeneratorSection}

Let us suppose that we have some solutions to the
syzygy equation~(\ref{GeneralSyzygy}), but not a complete
set.  Can we generate additional solutions?  

\def\PermS{\textrm{Perm}}
\def\mopen{[\mskip -2mu[}
\def\mclose{]\mskip -2mu]}
Define the following operation on a vector $v[\sigma](s_{ij})$
of polynomials in the $s_{ij}$, where the vector is indexed by
the permutation $\sigma$,
\begin{equation}
\PermS[\rho]\bigl(v[\sigma](s_{ij})\bigr)
=\omega_{\rho^{-1}\circ\sigma}
\,v[C(\rho^{-1}\circ\sigma)](s_{\rho(ij)})\,.
\label{GeneratingSolutions}
\end{equation}
Here $\omega_\sigma$ is the sign (if any) 
that arises from putting $\sigma$ into canonical form,
and $C$ is the combination of cyclic transformations and
possibly a reflection required to put the permutation into
canonical form.
In \textsl{Mathematica\/} notation, given the fixed set of momentum labels $A$, the action of a permutation
$\rho$ on a subset $a\subset A$ is given by,
\begin{equation}
\rho(a) = a \mathsf{\,/. Thread}[A \to A\,\mopen\mskip 2mu\rho\mskip 2mu\mclose]
\end{equation}
while the composition of permutations 
($\rho_1\circ\rho_2(a)=\rho_1(\rho_2(a))$, which in 
\textsl{Mathematica} notation is
\textsf{PermutationProduct[$\rho_2$,$\rho_1$]}) is given by,
\begin{equation}
\rho_1\circ\rho_2 = 
\rho_1\,\mopen\mskip 2mu \rho_2 \mskip 2mu \mclose\,.
\end{equation}
Then, 
\begin{equation}
\PermS[\rho](\PolyId{r}{n}[\sigma])
\end{equation}
is also a solution.

\section{Syzygies of All-Plus Amplitudes}
\label{AllPlusSection}

\def\Perm{{\cal V}}
\def\Syz{{\cal S}}
\def\SyzAP{\Syz^{(+)}}
Let us apply the techniques discussed in previous sections
to the leading-color
all-plus amplitude~\cite{AllPlus} in pure Yang--Mills theory.
For the four-point
amplitude, we have three permutations organized as a vector,
\begin{equation}
\permset4 = \bigl((1\,2\,3\,4)\,,
(1\,2\,4\,3)\,,
(1\,3\,2\,4)\bigr)\,.
\label{FourPointPermutations}
\end{equation}
As expected from the general arguments given 
in \sect{GeneralSyzygySection}, 
we have two basis syzygies, both of degree zero,
\begin{equation}
\SyzAP_4 = \langle(1,\,-1,\,0)\,,
(1,\,0,\,-1)\rangle\,.
\label{FourPointSyzygies}
\end{equation}
These provide the initial conditions for an inductive
calculation (or equivalently the stopping conditions for
a recursive one).  The order of components in each element
of $\SyzAP_4$ follows \eqn{FourPointPermutations},
\begin{equation}
v\cdot A(\permset{n}) = \sum_{j=1}^{|\permset{n}|} 
v_j\,A((\permset{n})_j)\,,
\end{equation}
for any syzygy $v\in \SyzAP_n$.

For the five-point amplitude, there are twelve permutations to
consider,
\begin{equation}
\begin{aligned}
\permset5 = \bigl(&(1\,2\,3\,4\,5)\,,
(1\,2\,3\,5\,4)\,,
(1\,2\,4\,3\,5)\,,
(1\,2\,4\,5\,3)\,,
(1\,2\,5\,3\,4)\,,
(1\,2\,5\,4\,3)\,,
\\&
(1\,3\,2\,4\,5)\,,
(1\,3\,2\,5\,4)\,,
(1\,3\,4\,2\,5)\,,
(1\,3\,5\,2\,4)\,,
(1\,4\,2\,3\,5)\,,
(1\,4\,3\,2\,5)\bigr)\,.
\end{aligned}
\label{FivePointPermutations}
\end{equation}
As expected from the general arguments 
in \sect{GeneralSyzygySection}, we have ten relations.
These turn out to be six degree-zero relations, as found
by BBDJS and DM, and four degree-two relations, as also
found by BBDJS.  They can be generated 
using \eqn{GeneratingSolutions} from just two relations,
one of degree zero, and one of degree two:
\begin{equation}
\begin{aligned}
&(0,\,0,\,0,\,0,\,0,\,0,\,1,\,1,\,-1,\,-1,\,1,\,-1)\,,
\\& 
(0,\,0,\,0,\,Q_1(1,2,3,4,5),\,Q_1(3,4,5,1,2),\,0,\,0,\,
Q_1(3,2,1,5,4),\, Q_1(5,1,2,3,4),\,0,\,
\\&\hphantom{(} Q_1(2,3,4,5,1),\,0)\,,
\end{aligned}
\end{equation}
where,
\begin{equation}
Q_1(i_1,\ldots,i_5) = s_{i_1i_2}\,s_{i_2i_3}
+s_{i_1i_5}\,s_{i_2i_3}+s_{i_1i_2}\,s_{i_3i_4}
-s_{i_1i_5}\,s_{i_3i_4}-s_{i_2i_3}\,s_{i_4i_5}
-s_{i_3i_4}\,s_{i_4i_5}\,.
\end{equation}
Although we have not searched exhaustively,
these generating solutions are likely to be minimal both
in number and complexity.  In generating the complete
set of syzygies from these two, it is important to use
all permutations of the five external legs, and not just
those in \eqn{FivePointPermutations}.  The complete set
of syzygies is given in the auxiliary \textsl{Mathematica\/}
file \textsf{syzygies5-ap.m}.

For the six-point amplitude, the 60 permutations are
given in the auxiliary file.  Here, the basic 
factorization equations
do not yield the lowest-degree form of all the degree-two
syzygies; but we can generate them from a small set of
vectors.  For this amplitude, we find 58 syzygies in total,
as expected from \sect{GeneralSyzygySection}.  We find
the 25 degree-zero syzygies found by BBDJS and DM, as
well as 33 additional degree-two syzygies (one more than
reported by BBDJS).  

The number of required generators
depends on whether we want to generate the degree-zero
relations directly with constant coefficients, or whether
we are content to generate a set of degree-two identities
whose linear combinations are multiples of the degree-zero ones.
In the latter case, one degree-two generator suffices; in the 
former case, we need two degree-zero generators.  The
latter-case degree-two generator is not the simplest one possible,
but it is the simplest single generator we have found.  (There
is a simpler generator that yields 15 relations.)  All these
generators, along with the complete set of relations, are
given in the auxiliary \textsl{Mathematica\/} file
\textsf{syzygies6-ap.m}.  The results given there are
unlikely to be in the simplest form possible.

In the case of the seven-point amplitude, we are able to
find a single degree-two generator using the numerical 
linear algebra method of section \ref{NumericalSyzygiesSection}.
Using all permutations of this generator 
in \eqn{GeneratingSolutions}
yields all 358 relations, including [invariant-dependent
multiples of] the degree zero ones.
We give this
generator and the complete set of relations in an auxiliary \textsl{Mathematica\/} file
\textsf{syzygies7-ap.m}.

\section{Syzygies of Tree-Level MHV Amplitudes}
\label{MHVSection}

The all-plus one-loop leading-color amplitude is special
in that the only complex factorization channels it has
are at zeros of $\spa{i}.j$ products.  There is another
amplitude that shares this property: the tree-level
MHV amplitude.  This means that its syzygies also satisfy
\eqn{PoleSyzygyEquation} and hence \eqn{ReducedSyzygyGeneral}.
The difference is in the
initial conditions; the four-point amplitude has only
one degree-zero relation, the photon decoupling identity,
\begin{equation}
(1,\,1,\,1)\,,
\end{equation}
while the second identity is the simplest BCJ
identity,
\begin{equation}
(0,\, s_{12},\, -s_{23})\,.
\end{equation}
\def\vsix{v^{(6)}}
\def\vseven{v^{(7)}}
The process of solving the equations is similar, however.
For the five-point amplitudes, we expect no additional relations
beyond the KK and BCJ ones, and indeed do not find any.  
For the
six-point amplitudes, we ultimately find four additional
degree-two relations beyond the KK and BCJ ones, for
a total of 58 relations (as also found by BBDJS).  
These relations are given
in the auxiliary \textsl{Mathematica\/} file 
\textsf{syzygies6-mhv.m}.  The degree-two relations
are generated by a simple vector $\vsix$, with
\begin{equation}
\begin{gathered}
\vsix_1=s_{16}s_{45},\quad \vsix_2=s_{15}(s_{14}+s_{46}),\quad \vsix_5=s_{14}s_{15},\quad \vsix_6=s_{14}(s_{15}+s_{45}),\quad \vsix_{17}=s_{12}s_{14}\,,
\end{gathered}
\end{equation}
and remaining components zero.  Amusingly, the same vector
generates all 58 independent relations, though not in their
simplest form.

\newcommand{\veight}{v^{(8)}}
We similarly find generators of degree-two relations for the seven- and eight-point amplitudes.
They are vectors $\vseven$ and $\veight$ with non-zero entries,
\begin{equation}
\begin{gathered}
	\vseven_7=s_{17} \left(s_{14}+s_{46}+s_{47}\right),\quad 
	\vseven_9=\vseven_{15}=s_{17} \left(s_{14}+s_{47}\right),\quad
	\vseven_{10}=\vseven_{16}=s_{14} s_{17}\\
	\vseven_{17}=s_{15}s_{47},\quad
	\vseven_{18}=s_{14} \left(s_{17}+s_{57}\right),\quad
	\vseven_{31}=s_{17} s_{24}\,,
\end{gathered}
\end{equation}
and
\begin{equation}
\begin{gathered}
	\veight_1=s_{67} \left(s_{15}+s_{56}+s_{57}+s_{58}\right),\quad 
	\veight_7=s_{67} \left(s_{15}+s_{57}+s_{58}\right),\quad
	\veight_9=s_{67} \left(s_{15}+s_{58}\right),\\
	\veight_{10}=\veight_{52}=\veight_{62}=\veight_{65}= s_{67}s_{15},\quad
	\veight_{49}=s_{67}\left(s_{15}+s_{57}+s_{58}\right)+s_{46} s_{57},\\
	\veight_{51}=s_{67} \left(s_{15}+s_{58}\right),\quad
	\veight_{55}=s_{47} s_{56},\quad
	\veight_{61}=s_{67} \left(s_{15}+s_{58}\right),\quad
	\veight_{265}= s_{67}s_{25}\,.
 \end{gathered}
\end{equation}
Just as in the six-point case, these vectors generate all respectively 358 and 2518 relations between the tree-level seven- and eight-point MHV partial amplitudes, including those of degree-zero and degree-one.

We provide all syzygies as well as the generators above for six-, seven- and eight-point amplitude relations in the ancillary files
\textsf{syzygies6-mhv.m}, \textsf{syzygies7-mhv.m}, and \textsf{syzygies8-mhv.m}.

\section{An Observation}
\label{BoseSymmetryObservation}

\def\Sym{\textrm{Sym}}
We can define a complete symmetrization operation on syzygies
of $n$-point amplitudes,
\begin{equation}
\Sym\bigl(v[\sigma](s_{ij})\bigr) = 
\sum_{\rho\in S_n} \PermS[\rho]\bigl(v[\sigma](s_{ij})\bigr)\,.
\label{SymmetrizationOperation}
\end{equation}
For the four- and five-point all-plus amplitudes, we
find that,
\begin{equation}
\Sym \,v = 0\,,
\end{equation}
for all syzygies.  This is no longer generically
true for the six-point
amplitudes in a representation with 25 degree-zero 
syzygies and 33 degree-two ones.  It is, however, possible to find
a representation in the six-point case where all
but one degree-zero syzygies vanish under this symmetrization operation, and where all degree-two syzygies vanish.  As
the vanishing under symmetrization is not invariant under
multiplication by a polynomial of the invariants, it is
also possible to find a non-minimal-degree representation where
all syzygies vanish under symmetrization.

\section{Conclusions}
\label{ConclusionsSection}

The internal structure of Yang--Mills
scattering amplitudes is expressed through linear
relations between color-ordered amplitudes.  
The best-known relations~\cite{Kleiss:1988ne}
between amplitudes with constant coefficients emerge
directly from Lie-algebra relations.  The work of
DM suggests that other degree-zero relations also have
a partly group-theoretic origin.
Additional relations have coefficients which are polynomial
in the momentum invariants.  The BCJ identities are
well-known examples of such relations.  Their origin
is still unclear, but it is beyond group theory.

We showed in \sect{GeneralSyzygySection} that such relations
must exist on general grounds, leaving two independent amplitudes
for five or more external particles (one independent
amplitude in the four-point case).  The general argument
does not force these relations to be of low polynomial
degree; but explicit forms obtained by BBDJS and DM, and
extended here, are of quadratic degree or lower for
the one-loop all-plus amplitude and the tree-level MHV
amplitude.  We have checked through $n=7$ for the all-plus
amplitude and $n=8$ for the MHV one that these exhaust
all relations required by the general argument.  We conjecture
that this statement holds to all $n$.

We presented two techniques for solving for the relations,
one numerical and one mostly analytic,
based on the factorization properties
of amplitudes.  The latter are presumably responsible for
the surprisingly low degree of the relations.  The factorization
equations we derived are special to amplitudes with only
three-point amplitudes in all factorization channels.
They can be extended to more general factorizations, a subject
we leave to future study.  The generality of the arguments
suggests that similar relations should also exist for 
rational terms in higher-loop amplitudes, and for the
rational coefficients of integral functions as well.
They may prove useful in reducing computational complexity
in numerical calculations, and in understanding new symmetry
structures underlying Yang--Mills theories.  A generalization
of the relations studied here also describes the redundancy
in Kawai--Lewellen--Tye kernels~\cite{Kawai:1985xq,KLTinQFT}
relating tree-level Yang--Mills and gravity amplitudes.

\begin{acknowledgments}
We thank CERN, the GGI, the KITP, and the Perimeter Institute for their hospitality
during phases of this work.  We thank Lance Dixon
and Anthony Morales for helpful discussions and comments,
and Roland Bittleston and Kevin Costello for helpful discussions.
This research was supported by the European Research
Council (ERC) under the European Union’s research and
innovation programme grant agreements ERC--AdG--885414 (`Ampl2Einstein');
in part by grant NSF PHY--2309135 to the Kavli 
Institute for Theoretical Physics (KITP); and in part
by a grant from the Simons Foundation (1034867, Dittrich).
Research at Perimeter Institute is supported in part by the Government of Canada through the Department of Innovation, Science and Economic Development and by the Province of Ontario through the Ministry of Colleges and Universities. 

\end{acknowledgments}

\FloatBarrier

\end{document}